\documentstyle[sprocl]{article}

\input epsf
\def\be{\begin{equation}}
\def\ee{\end{equation}}
\def\bea{\begin{eqnarray}}
\def\eea{\end{eqnarray}}

\begin{document}
\title {PROBING THE STRUCTURE OF NUCLEONS IN THE RESONANCE REGION}

\author{VOLKER D. BURKERT}

\address{Thomas Jefferson National Accelerator Facility, \\ 
12000 Jefferson Avenue, 
 Newport News, Virginia, USA\\
email: burkert@jlab.org}


\maketitle\abstracts{Status, open questions, and future prospects of the physics of excited
nucleons, and what they tell us about the internal nucleon structure are discussed.}

\section{Introduction}
Electromagnetic production of hadrons may be characterized according to 
distance and time scales (or momentum and energy transfer) 
probed in the interaction. This is illustrated with the 
three regions in Figure 1. For simplicity I have omitted the time scale. 
At large distances mesons and nucleons are the 
relevant degrees of freedom. Due to the limited spatial resolution 
of the probe we study peripheral properties of nucleons
near threshold for pion production. Chiral perturbation theory describes many 
of these processes and has a direct link to QCD via (broken) chiral symmetry. 
 At short distances (and short time scales), the coupling involves elementary quark and 
gluon fields, governed  by perturbative QCD, and we map out parton distributions in the nucleon.  
At intermediate distances, quarks and gluons are 
relevant, however,  confinement is important, and they appear as constituent 
quarks and glue. We study interactions between these constituents via their excitation
 spectra and wave functions. This is the region where the connection to the fundamentals 
of QCD remains poorly established, and where JLab experiments currently have their 
biggest impact.
These regions are not strictly separated from each other but 
overlap, and the hope is that due to this overlap the nucleon structure 
may eventually be described in a more unified approach, based on fundamental
 theory, from small to large distances.
Because the electromagnetic probe is well understood, 
it is well suited to provide the data for such an endeavor. 

\begin{figure}[htbp]
\epsfysize=15.0truecm
\epsfbox{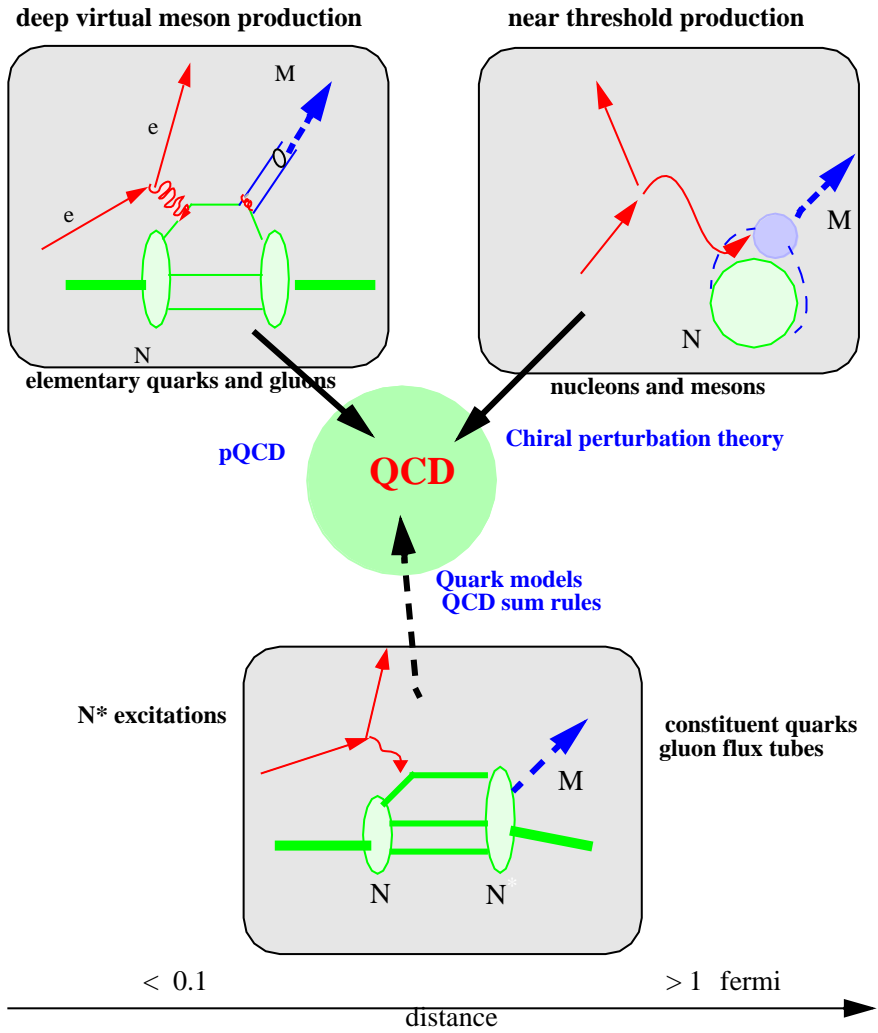}
\caption {Exclusive meson electroproduction. A subdivision in
distance scales is used to illustrate three kinematic regions and their 
respective (effective) degrees of freedom.}
\end{figure}

\subsection{ Structure of the Nucleon at Intermediate Distances - 
Open Problems}

QCD has not been solved for processes at intermediate distance scales. A direct 
consequence is that the internal structure of nucleons is 
poorly known. On the other hand, theorists are often 
not challenged due to the lack of high quality data in many areas. 
The following are areas where the lack of high quality data is most noticeable, and
where data from CLAS are expected to contribute significantly in the future. Some of
the first results will be presented at this conference.

\begin{itemize}

\item{The form factors of the neutron $G_{En}$, $G_{Mn}$ are poorly known. 
This means that the charge and magnetization distribution of the basic 
building blocks of matter in the visible universe is virtually unknown.} 

\item{To understand the internal nucleon structure, 
it is not sufficient to measure the charge and 
current distribution in the ground state, we need to study the full excitation 
spectrum as well as the continuum. 
Few transitions to excited states have been studied well, and many
states are missing from the spectrum as predicted by our most accepted models.} 

\item {The role of the glue in the baryon excitation spectrum is completely unknown, 
although gluonic excitations of the nucleon are expected to be produced 
copiously \cite{isgur}, and predictions of hybrid baryon masses and quantum numbers 
are available.}  

\item{The nucleon spin structure has been explored for more than two decades 
at high energies in laboratories such as CERN, SLAC, and DESY. 
The confinement regime and the transition between these regimes have not been 
explored at all.}

\item {The long-known connection between the deep inelastic regime and the 
regime of confinement (parton-hadron duality) \cite{blogil} 
remained virtually unexplored in its potential implications 
for theoretical developments.}

\end{itemize}

{\sl Carrying out an experimental program that will address these questions 
has become feasible due to the 
availability of CW electron accelerators, modern detector instrumentation with 
high speed data transfer techniques, the routine availability of spin
polarization in beam and targets, and recoil polarimetry.}

\section{Excitation of Baryon Resonances}

A large effort is being extended to the study of excited states of the nucleon. 
The  transition form factors contain information 
on the spin structure of the transition and the wave function of the 
excited state. We test predictions of baryon structure models and of strong interaction 
QCD.
Another aspect is the search for, so far, unobserved states which 
are missing from the spectrum but are predicted 
by QCD inspired quark models \cite{isgur2}. Also, are there other than $|Q^3>$ states? 
Gluonic excitations of the nucleon, i.e. $|Q^3G>$ states
may be be copious \cite{isgur}, and some resonances may be 
``molecules'' of baryons and mesons 
$|Q^3Q\bar Q>$.
Searching for at least some of these states is important to clarify the intrinsic 
quark-gluon structure of baryons and 
the role played by the glue and mesons in hadron spectroscopy and structure. 
Photo- and electroproduction of mesons are important tools in these studies 
as they probe the internal 
structure of hadronic systems.
The scope of the $N^*$ program \cite{nstar} at JLAB includes
measurement of many of the possible decay channels of resonances in a 
large kinematical range.

\subsection{\bf The $\gamma N \Delta$ transition.}

The lowest excitation of the nucleon is the $\Delta(1232)$ ground state. The 
electromagnetic excitation is due dominantly to a quark spin flip corresponding
to a magnetic dipole transition. This contribution is fairly well known 
up to quite large $Q^2$.
The current interest is in probing the electric and scalar quadrupole
transitions which are predicted to be sensitive to a possible deformation of 
the nucleon or the $\Delta(1232)$.
Contributions at the few percent level may result from interactions with 
the pion cloud \cite{yang} at large and intermediate distances, and from one-gluon-exchange 
at small distances. 
An intriguing prediction is that in the hard scattering limit the 
electric quadrupole contribution should be equal in strength to the 
magnetic dipole contribution \cite{carlson}. An analysis \cite{burelou} 
of earlier DESY data
found small nonzero values for the ratio $E_{1+}/M_{1+}$ at $Q^2 = 3.2GeV^2$, 
showing that the
asymptotic QCD prediction is far away from the data. 
 
An experiment at JLAB Hall C \cite{frolov,stoler} measured  
$p\pi^o$ production in the 
$\Delta(1232)$ 
region at high momentum transfer, and found values for
$E_{1+}/M_{1+} \approx -0.02$ at $Q^2 = 4~GeV^2$.

Preliminary data from CLAS \cite{lcsmith}
indicate negative values at small $Q^2$ 
with a trend towards more positive values at higher $Q^2$. Much more data 
are expected in the near future from CLAS. 

\subsection {\bf What is so special about the Roper Resonance ?}

The lowest positive parity $N^*$ resonance is the $P_{11}(1440)$. 
Its internal structure has been subject of intensive debate in 
recent years. It is clearly visible in $\pi N$ scattering, as well as in 
pion photoproduction. However, its transition strength drops rapidly with $Q^2$ in 
electroproduction \cite{gerhardt}, and its longitudinal coupling
is weak. Neither of these properties is well described in 
quark models, which assigns the state to a radial excitation of the 
nucleon. It has been proposed that the observed 
``electro-quenching'' behavior 
is well described if the state is assigned a large gluonic component 
\cite{libuli}. While recent flux tube model calculations \cite{page} 
give higher masses to hybrid baryons than previous estimates in the bag model 
\cite{gohaka} and QCD sum rules\cite{kissl}, the Roper could still have a 
substantial gluonic component due to mixing with higher 
mass states \cite{capstick}.
Other models have been put forward recently
that include meson degrees of freedom \cite{cato}, or 
 describe the state as a molecule of the nucleon and 
the $\sigma$ pseudo-particle \cite{krewald}. The soft formfactor
suggests an extended or loosely bound system. 
Studying these transitions in electroexcitation allows us to probe 
the internal structure and help reveal the nature of this state.
CLAS $\pi^o$ and $\pi^+$ electroproduction data are currently being 
analyzed to measure the transition amplitudes in a large range of $Q^2$.

\subsection {\bf Higher mass resonances}

The inclusive spectrum shows only 3 or 4 enhancements; however, 
more than 20 states are known in the mass region up to 2 GeV. 
 By measuring the electromagnetic transition of many of these 
 states we obtain a  more complete picture of the 
nucleon structure. 

The approximate SU(6) symmetry of the non-relativistic 
symmetric quark model predicts relationships between various 
states. For example, in the single-quark-transition model (SQTM) 
only one quark
participates in the interaction. It predicts transition amplitudes 
for a large number of states based on a few measured amplitudes \cite{hey}.
The current situation is shown in Figure 5, where the SQTM amplitudes for 
the transition to the $L_{3q}=1$ $SU(6)\otimes O(3)$ multiplet have been 
extracted from
the measured amplitudes for $S_{11}(1535)$ and $D_{13}(1520)$. Predictions 
for other
states belonging to the same multiplet are shown in the other panels. 
The lack of accurate data for most other resonances prevents a sensitive test of 
even the simple SQTM.

\begin{figure}[htbp]
\epsfysize=15.0truecm
\epsfbox{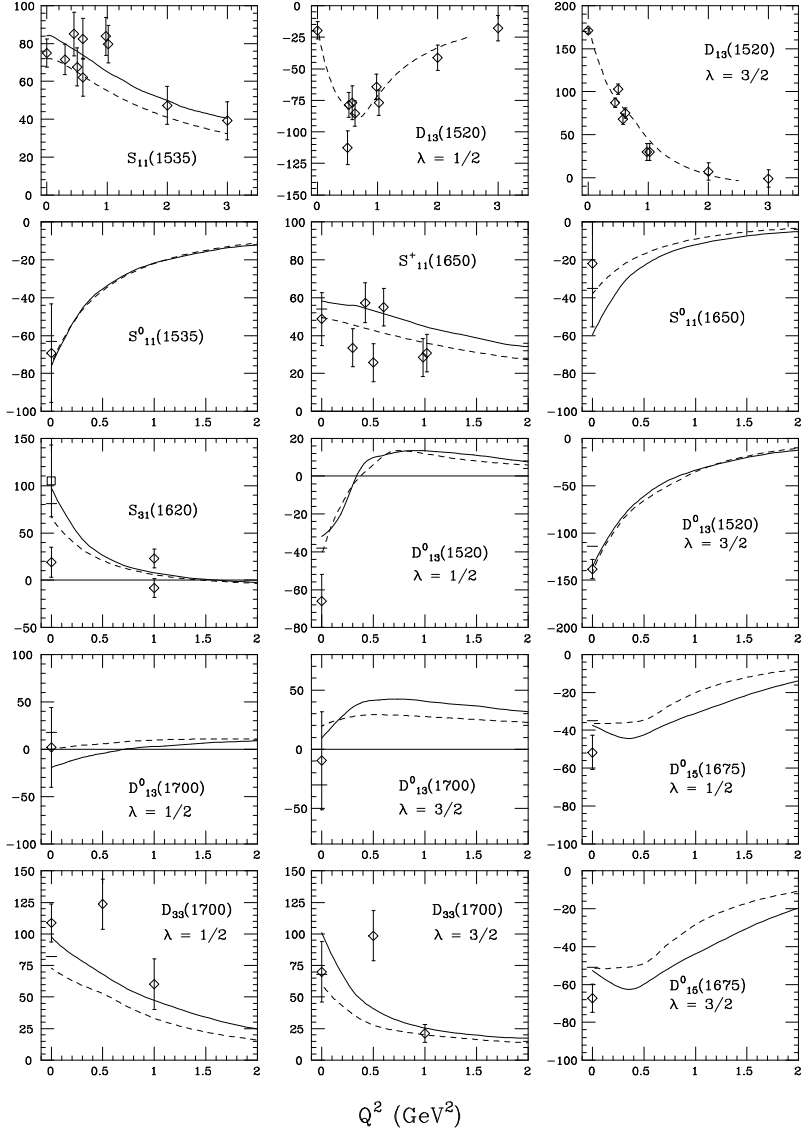}
\hsize=11cm
\caption{\small Single Quark Transition Model predictions for states 
belonging to the $SU(6) \otimes O(3)$ multiplet, discussed in the text.}
\end{figure}

The goal of the experimental N* program at JLAB with the CLAS detector is to 
provide data in the entire resonance region, by measuring many
channels in a large kinematic range, including various polarization observables. 
The yields of several channels recorded simultaneously are shown in Figure 3
and Figure 4. Resonance excitations seem to be present in all channels.
The figures also illustrate how the various channels have sensitivity 
to different resonance excitations. For example, the $\Delta^{++}\pi^-$ channel 
clearly shows resonance excitation near 1720 MeV while single pion 
production is more sensitive to a resonance near 1680 MeV 
\cite{ripani}. The $p\omega$ channel seems to show resonance excitation near threshold, 
similar to the $p\eta$ channel. No resonance has been observed in this channel 
so far. For the first time,  $n\pi^+$ electroproduction has been measured 
throughout the resonance region, and in a large angle and $Q^2$ range. 

\begin{figure}[t]
\epsfysize=8.5truecm
\epsfxsize=7.5truecm
\hspace{2cm}
\epsfbox{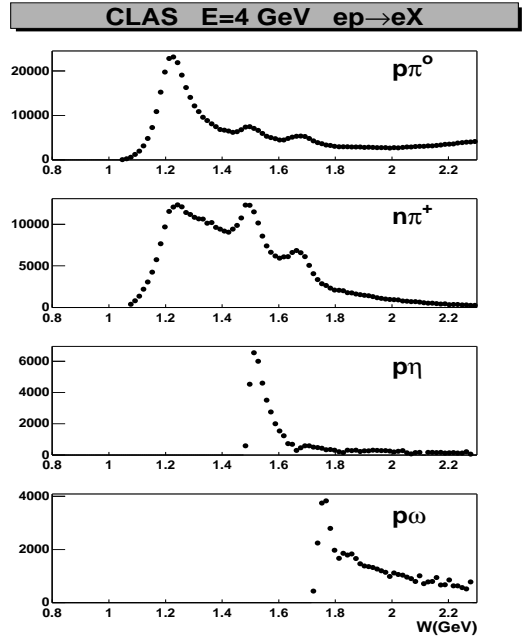}
\caption{\small Yields for various channels measured with CLAS at JLAB. 
The statistical error bars are 
smaller than the data points.}
\end{figure}

Figure 4 illustrates  the vast improvement in data volume for the 
$\Delta^{++}\pi^-$ channel. 
The top panel shows DESY data taken more than 20 years ago.  The other 
two panels show samples of the data taken so far with CLAS. At higher $Q^2$, 
resonance structures, not seen before in this channel, are revealed.

\subsection {\bf Missing quark model states} 

These are states predicted in the $|Q^3>$ model to populate the 
mass region around 2 GeV.  However, they have
not been seen in $\pi N$ elastic scattering, our main source of information 
on the nucleon excitation spectrum. 

It is important to search for at least some of these states since
their absence from the spectrum would be evidence that either 
SU(6) symmetry is strongly broken in baryon spectroscopy, or 
quark model calculations of electromagnetic or hadronic couplings 
are unreliable.  
How do we search for these states?

Channels which are predicted to couple strongly to these states are
$N(\rho, \omega)$ or $\Delta\pi$ \cite{ripani}. 
Some may also couple to $KY$ or $p\eta^{\prime}$ \cite{caprob}. 

Figure 5 shows preliminary data from CLAS in $\omega$ production 
on protons. The 
process is expected to be dominated by $\pi^o$ 
exchange with strong peaking at forward $\omega$ angles, or low t,  
and a monotonic fall-off at large t. 
The data show clear deviations 
from the smooth fall-off for the W range near 1.9 GeV, where some of the 
``missing'' resonances are predicted, in comparison with the high W region. 

Although indications for resonance production are strong\cite{funsten},
analysis of more data and a full partial wave study are needed before 
definite conclusions may be drawn. 

\begin{figure}[t]
\epsfysize=8.0truecm
\epsfxsize=7.0truecm
\hspace{2cm}
\epsfbox{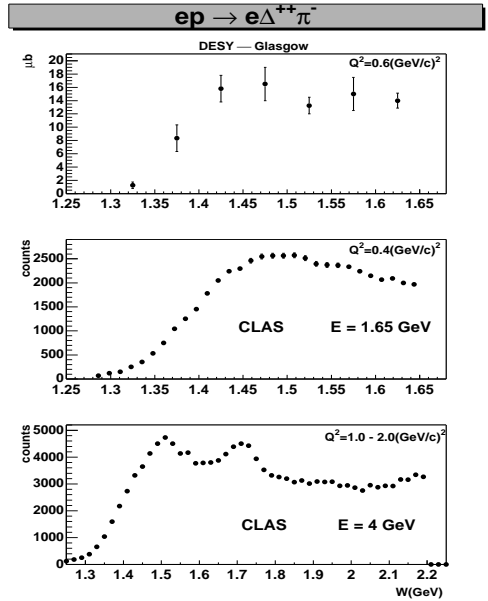}
\caption{\small Yields for the channel $\Delta^{++}\pi^-$ measured with CLAS at 
different $Q^2$ compared to previous data from DESY.}
\end{figure}

CLAS has collected ~$3\cdot 10^5$ $p\eta^{\prime}$ events
in photoproduction. Production of $\eta^{\prime}$ has 
also been observed in electron scattering for the first time with CLAS. 
This channel may provide a new tool in the search for missing states
as well \cite{ritchie}.
The quark model predicts two resonances in this mass range 
with significant coupling to the $N\eta^{\prime}$ channel \cite{caprob}. 

$K\Lambda$ or $K\Sigma$ production may yet be another source of information on resonant
 states. Previous data show some evidence for resonance production in these 
 channels.
New data with much higher statistics are being accumulated with the CLAS
detector, both in photo- and electroproduction \cite{mestayer}. Analysis
of the $\Lambda$ polarisation possible in CLAS provides additional information
sensitive to resonance excitations.

\begin{figure}[t]
\begin{minipage}{0.49\textwidth}
\epsfysize=6.0cm
\epsfxsize=5.5cm
\epsfbox{fig8_e99.epsi}
\hsize=5.6truecm
\caption{Electroproduction of $\omega$ mesons for different W bins. 
The deviation 
of the $\cos\theta$ -distribution from a smooth fall-off for the 
low W bin suggests significant 
s-channel resonance production.}
\end{minipage}
\begin{minipage}{0.49\textwidth}
\epsfysize=6.0cm
\epsfxsize=5.5cm
\epsfbox{fig9_e99.epsi}
\hsize=5.6truecm
\hspace{0.3truecm}
\caption{Ratio of resonance excitations as observed 
and predicted from 
deep inelastic processes using quark-hadron duality {\protect\cite{keppel}}.}
\end{minipage}
\end{figure}

\section{The Nucleon Spin Structure - from Small to Large Distances} 

The internal spin structure of the nucleon has been of central interest ever since the 
EMC experiment 
found that at small distances the quarks carry only a fraction of the
nucleon spin. 
Going from small to large distances the quarks get dressed with gluons 
and $q\bar q$ pairs 
and acquire more and more of the nucleon spin. How is this process evolving 
with the distance scale? 
At the two extreme kinematic regions we have two fundamental sum rules. The 
Bjorken sum rule (Bj-SR) which holds in the asymptotic 
limit is usually written for the proton-neutron difference as 
$$ \Gamma_1^{pn} = \int{g_1(x)dx} = {g_A \over 6}~~.$$ At the finite $Q^2$ where 
experiments are performed, 
QCD corrections have been calculated, and there is good agreement between theory and experiment at 
$Q^2> 2~ GeV^2$.
At the other end, at $Q^2 = 0$, the Gerasimov Drell-Hearn 
sum rule (GDH-SR) is expected to hold: 
$$I_{GDH} = {M^2\over 8\pi^2\alpha}\int {{\sigma_{1/2}(\nu)-\sigma_{3/2}(\nu)}\over \nu}d\nu = 
-{1\over 4}\kappa^2 ~~ .$$
The integral for the difference in helicity 1/2 and helicity 3/2 total 
absorption cross sections 
 is taken over the entire inelastic energy regime. The quantity $\kappa$ is the anomalous magnetic moment of the target.

One connection between these regions is given by the constraint due to the GDH-SR 
- it defines the slope of the Bjorken integral
 ($\Gamma_1(Q^2) = \int g_1(x,Q^2)dx$) 
at $Q^2=0$: $$ I^{pn}_{GDH}(Q^2 \rightarrow 0) = 2{M^2\over Q^2}  \Gamma_1^{pn} (Q^2 \rightarrow 0) $$ 

\noindent 
Phenomenological models \cite{buriof1,buriof2,soffer} have been 
proposed to extend the GDH integral for the 
proton and neutron to finite $Q^2$ and connect it to the deep inelastic 
regime. The low $Q^2$ data from SLAC experiment E143 \cite{abe} 
are in good agreement with predictions \cite{buriof2} if nucleon resonances are taken into 
account explicitly (Figure 7).

\noindent
An important question is whether we can go beyond models and describe the 
transition from the Bj-SR to the  GDH-SR within 
fundamental theory, i.e. QCD. X. Ji and collaborators \cite{ji1} have recently 
generalized the sum rule integral to include finite $Q^2$. The question is if
the right-hand side of the integral can be predicted by theory.
Chiral perturbation theory has been proposed \cite{ji1,meissner}
to provide an extension of the sum rule for proton and neutron 
to finite $Q^2$.
If we look at this problem for the proton or neutron separately, we find that 
the sum rule is nearly saturated by low-lying resonances \cite{burkert1,tiator} with
the largest contributions coming from the excitation of the $\Delta(1232)$. This
is still correct if the GDH-SR is generalized to finite, but not too large $Q^2$ 
\cite{burkert1,ma}. This will make it very difficult to describe the sum rule 
within chiral perturbation theory\cite{jikaos}  at any, but the smallest 
$Q^2$. The situation looks more promising if we take the proton-neutron 
difference \cite{burkert3}. The dominant contribution from the $\Delta(1232)$ 
is absent, and other resonance contributions are reduced as well indicating 
that the GDH-SR may be evolved to $Q^2 \approx 0.2$ GeV$^2$ if higher orders 
are taken into account. An important question in this connection is: how low 
in $Q^2$ may the Bjorken-SR be evolved using higher twist expansion?
Recent estimates  \cite{jikaos} suggest these techniques may be valid as low 
as $Q^2 = 0.5$ GeV$^2$, leaving a gap from $Q^2 = 0.2 - 0.5$ 
that need to be bridged, perhaps utilizing lattice QCD. 

\begin{figure}[t]
\epsfysize=7.5truecm
\epsfxsize=7.5truecm
\hspace{1.5cm}
\epsfbox{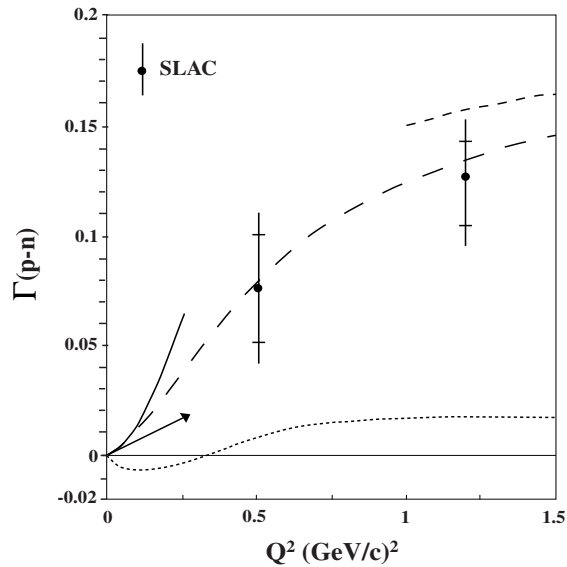}
\caption{The Bjorken integral of the polarized structure 
 function $g_1(x,Q^2)$ of the proton and neutron. The solid line is 
the $\chi$PT 
 prediction \protect\cite{burkert3,jikaos}. 
Short dashed line is the Bjorken 
 sum rule with pQCD corrections to O($\alpha_s^3$). 
 The long dashed line is a prediction including s-channel 
 resonances \protect\cite{buriof2}. 
 The dotted line shows s-channel resonance 
contributions \protect\cite{burkert1}.
 The solid arrow indicates the slope given by the GDH sum rule.}
\end{figure}

The connection of deep inelastic regime, higher twist,
and resonance contributions for the proton spin structure 
function $g_1(x,Q^2)$ has recently 
been studied by Weise and collaborators \cite{weise}. They estimate 
sizeable resonance contributions even at high $Q^2$.  

These efforts are of utmost importance since, if successful, 
{\it it would mark the first time that 
hadronic structure is described by fundamental theory in the entire kinematic
regime, from small to large distances, a worthwhile goal!}  

Experiments have been carried out at JLAB \cite{minehart,kuhtai,jiang} on $NH_3$, $ND_3$, 
and $^3He$ 
targets to extract the $Q^2$ evolution of the generalized GDH integral 
for protons and neutrons
in the low $Q^2$ range,  $Q^2 = 0.1 - 2.0~GeV^2$, and 
from the elastic to the deep inelastic regime. Currently, only two data points with large errors exist for 
$Q^2 < 2~ GeV^2$. Because of the current limitations in machine energy to 6 GeV, some 
extrapolation will be needed to determine the full integral, especially at the 
larger $Q^2$ values. 
First results from the JLAB experiments are expected in the year 2000 
\cite{minehart}.

\section{\bf Connecting Constituent Quarks and Valence Quarks}

I began my talk by expressing the expectation that we may eventually 
arrive at a unified description of hadronic structure from small to 
large distances. If such description is possible then there should 
be obvious connections in the data between these regimes. I have 
already mentioned 
the evolution of the Gerasimov-Drell-Hearn and Bjorken sum rules 
as one area where 
such connections are visible in the data, and where 
significant progress seems likely in the near future. 

Inclusive electron scattering 
is another area where strong connections have been observed by 
Bloom and Gilman \cite{blogil}. They noted that the scaling curves from 
the deep inelastic cross section also describe the average 
inclusive cross sections in the resonance region if a scaling variable is 
chosen that takes into account target mass effects. Until recently, this intriguing
observation was little utilized. 
A high precision inclusive ep scattering experiment \cite{keppel} at JLAB helped rekindle
the interest in this aspect of hadron physics. 
Remarkably, resonance excitations of the nucleon can be 
predicted approximately 
from inclusive deep inelastic scattering data. 
Figure 6 shows the ratio of measured integrals over resonance 
regions, and predictions using deep inelastic data only. The agreement is 
surprisingly good, though not perfect.
It remains to be seen if this intriguing observation can be translated into 
the development of new theory approaches to resonance physics. 
Studies of exclusive channels, as well as of polarized structure functions in
the resonance region may contribute towards a deeper  
theoretical understanding
of this observation.

\section{Outlook}

The ongoing experimental effort at Jefferson Lab will provide the community 
with a wealth of 
data in the first decade of this millennium to address many open problems 
in hadronic structure in the resonance region and 
at intermediate distances.

{\it The experimental effort must be accompanied by a significant 
theoretical effort to translate this into real progress in our 
understanding of the complex regime of strong interaction physics}. 

The nucleon resonance region
is of special interest as it represents a region where different
degrees of freedom, from hadronic, to constituent quarks, to valence quarks
overlap. On the one hand this provides a challenge to theory, 
but on the other hand an opportunity, as only under such circumstances 
is there a realistic chance for a unified description of hadron structure 
from short to large distances.

\end{document}